\documentclass[journal=nalefd,manuscript=letter]{achemso}
\usepackage[version=3]{mhchem} 
\usepackage{xcolor}
\usepackage[T1]{fontenc}
\mciteErrorOnUnknownfalse

\author{Ik Kyeong Jin}
\affiliation[UNSW Physics]
{School of Physics, The University of New South Wales, Sydney NSW 2052, Australia}
\alsoaffiliation[RIKEN]
{Present address: RIKEN, 2-1 Hirosawa, Wako, Saitama 351-0198, Japan}
\alsoaffiliation[Equal]{These authors contributed equally to this work.}

\author{Joseph Hillier}
\affiliation[UNSW Physics]
{School of Physics, The University of New South Wales, Sydney NSW 2052, Australia}
\email{Joe.Hillier@unsw.edu.au}
\alsoaffiliation[Equal]{These authors contributed equally to this work.}

\author{Scott D. Liles}

\author{Zhanning Wang}

\author{Aaquib Shamim}
\affiliation[UNSW Physics]
{School of Physics, The University of New South Wales, Sydney NSW 2052, Australia}

\author{Isaac Vorreiter}
\affiliation[UNSW Physics]
{School of Physics, The University of New South Wales, Sydney NSW 2052, Australia}

\author{Ruoyu Li}
\affiliation[IMEC]
{IMEC, Remisebosweg 1, B-3001 Leuven, Belgium}
\author{Clement Godfrin}
\affiliation[IMEC]
{IMEC, Remisebosweg 1, B-3001 Leuven, Belgium}
\author{Stefan Kubicek}
\affiliation[IMEC]
{IMEC, Remisebosweg 1, B-3001 Leuven, Belgium}
\author{Kristiaan De Greve}
\affiliation[IMEC]
{IMEC, Remisebosweg 1, B-3001 Leuven, Belgium}

\author{Dimitrie Culcer}
\affiliation[UNSW Physics]
{School of Physics, The University of New South Wales, Sydney NSW 2052, Australia}

\author{Alexander R. Hamilton}
\affiliation[UNSW Physics]
{School of Physics, The University of New South Wales, Sydney NSW 2052, Australia}
\title{Probing g-tensor reproducibility and spin-orbit effects in planar silicon hole quantum dots}

\abbreviations{}

\begin{document}

\begin{abstract}

In this work, we probe the sensitivity of hole-spin properties to hole occupation number
in a planar silicon double-quantum dot device fabricated on a 300 mm integrated platform. Using DC transport measurements, we investigate the g-tensor and spin-relaxation induced leakage current within the Pauli spin-blockade regime as a function of magnetic-field orientation at three different hole occupation numbers.
We find the g-tensor and spin-leakage current to be highly anisotropic due to light-hole/heavy-hole mixing and spin-orbit mixing, but discover the anisotropies to be relatively insensitive to the dot hole number. 
Furthermore, we extract the dominant inter-dot spin-orbit coupling mechanism as surface Dresselhaus, with an in-plane orientation parallel to transport and  magnitude $\boldsymbol{t_{SO}}$ $\approx$ 300 neV. Finally, we observe a strong correlation between the g-factor difference ($\delta$$\boldsymbol{g}$) between each dot and the spin-leakage current anisotropy, as a result of $\delta$$\boldsymbol{g}$ providing an additional spin-relaxation pathway. 
Our findings indicate that hole-spin devices are not as sensitive to precise operating conditions as anticipated. This has important implications optimizing spin control and readout based on magnetic-field direction, together with tuning large arrays of QDs as spin-qubits.
\end{abstract}

Spin-qubits in semiconductor dots provide an industry-compatible platform to achieve future up-scaling through high packing densities \cite{Rotta2017-dc,vinet_towards_2018,doi:10.1126/science.abb2823}.
Numerous breakthroughs have been demonstrated with electron spin-qubits in silicon dots, including record gate fidelities  ($>$99.95 \%), high temperature operation and AC electric-field control for spin-state manipulation \cite{lim_electrostatically_2009,fogarty_integrated_2018,Huang2024-hv,yang_operation_2020,gilbert_-demand_2023,Camenzind2022-fy}. Recent interest in hole spins has been sparked by their intrinsic strong spin-orbit interaction (SOI) which enables fast all-electric control, the absence of a valley degree of freedom and a weak hyperfine interaction. \cite{de_greve_ultrafast_2011,lawrie_spin_2020,bulaev_electric_2007,watzinger_germanium_2018, hendrickx_fast_2020, Camenzind2022-fy, 
maurand_cmos_2016, hendrickx_four-qubit_2021, geyer_two-qubit_2022, bosco_phase_2023}.
These advantages have permitted a plethora of hole-spin qubit advancements, such as ultra-fast singlet-triplet operation ($>$400 MHz), tunable spin-orbit effects and coherent shuttling across large arrays \cite{liles2023singlettripletholespinqubitmos, froning_ultrafast_2021,Wang:2024isl}.
\\
Pauli spin-blockade is a commonly used technique for spin-readout within a double-dot, but in hole systems it is heavily influenced by the SOI \cite{nadj-perge_spinorbit_2010,Hillier2021-ns}. This results in unwanted spin-rotations and reduced fidelity when holes tunnel between dots. The strong SOI also gives rise to variations in the hole g-tensor on each dot due to differences in the local electronic environment, such as the dot hole number, local strain and gate biases \cite{ares_nature_2013,srinivasan_electrical_2016,bogan_consequences_2017,liles_electrical_2021}. While a g-factor difference ($\delta$$\boldsymbol{g}$) between dots is useful for driving rapid spin control in singlet-triplet qubits (which rely on differences in the Zeeman-splitting to drive qubit rotations) it can also lead to performance and scalability issues \cite{brauns_electric-field_2016,bogan_consequences_2017,jirovec_singlet-triplet_2021}. For example, it is not clear if the spin properties for one particular dot occupation remain the same when the number of holes is varied.
\\
Probing the SOI mechanism using spin-blockade based qubit readout via anisotropy of both the spin-orbit coupling ($\boldsymbol{t_{SO}}$) and hole g-tensor is also important for improving control and fidelity when scaling to larger numbers of qubits \cite{nadj-perge_disentangling_2010,nadj-perge_spectroscopy_2012,Wang2018-cy}. This is because, while a strong SOI enhances the coupling to more effectively drive the qubit spin state, it also couples the qubit to electrical noise, reducing the coherence time. Unfortunately, $\boldsymbol{t_{SO}}$ is a difficult parameter to measure \cite{harvey-collard_spin-orbit_2019,tanttu_controlling_2019}. One approach is to extract $\boldsymbol{t_{SO}}$ from singlet-triplet level anti-crossings, but this is technically challenging since it requires fast pulsed gate operations \cite{shevchenko_landauzenerstuckelberg_2010,higginbotham_hole_2014}. A simpler method to evaluate $\boldsymbol{t_{SO}}$ and hole g-tensor anisotropy would be highly desirable.
\\
Here, we take advantage of simple DC transport measurements of spin-leakage current ($I_{leak}$) to characterise the dominant $\boldsymbol{t_{SO}}$ mechanism, together with spin-resonance to probe the anisotropic g-tensor of a silicon hole double-dot at three different dot occupations, without fast pulses or time resolved measurements. $\boldsymbol{t_{SO}}$ is extracted by monitoring the 3D magnetic-field dependence of $I_{leak}$ within the Pauli spin-blocked regime. We utilize continuous-wave electric-dipole spin-resonance (EDSR) to extract the dot g-tensor ($\boldsymbol{g_1}$) in all planes and dot occupations as well as $\delta$$\boldsymbol{g}$ in-plane. Our results allow comparisons between the dependence of $I_{leak}$, $\boldsymbol{t_{SO}}$ and hole g-tensor anisotropy on the magnetic-field orientation when varying the hole number, to provide key information on the reproducibility and robustness of the spin properties for a path towards scalable hole-spin qubit architectures.

\newpage
\begin{figure} 
    \includegraphics{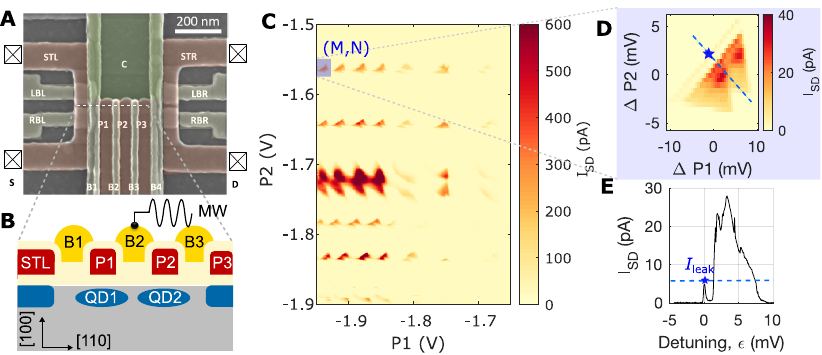}
	\caption{\label{fig:deviceSchmatics}} (\textbf{A}) SEM image of our planar Si-MOS dot device gate layout. (\textbf{B}) Device cross-section, where QD$_1$ and QD$_2$ are formed under plungers P1 and P2, with tunnelling barriers B1, B2 and B3 defining the potential longitudinally, and confinement gate C laterally. Holes are loaded from the STL reservoir and tunnel through the double-dot to the STR/P3 2D hole-gas along the [110] direction. (\textbf{C}) A charge stability map with bias triangles to confirm a well-defined double-dot structure. A periodic array of bias triangles enables a systematic study of different charge occupations. (\textbf{D}) An enlargement of a measured bias triangle as highlighted in Figure \ref{fig:deviceSchmatics}\textbf{C}. (\textbf{E}) A line-cut of detuning vs $I_{SD}$, where transport is measured along the detuning axis given by the blue dashed line in Figure \ref{fig:deviceSchmatics}\textbf{D}. A finite leakage current, $I_{leak}$, (marked by a blue star on the graph) is measured at the base of the bias triangle due to singlet-triplet mixing within the Pauli spin-blockade regime. 
\end{figure}

Figure \ref{fig:deviceSchmatics}\textbf{A} shows an SEM image of the planar silicon metal-oxide-semiconductor device gate layout, fabricated using a flexible 300 mm process with electron beam lithography for patterning the pitch-critical areas and optical lithography for defining larger features\cite{li_flexible_2020}. The device cross-section is displayed in Figure \ref{fig:deviceSchmatics}\textbf{B}, where the dots (QD$_1$ and QD$_2$) are vertically defined along the [100] growth direction and hole transport ($I_{SD}$) occurs in the [110] direction.  The dots are created under plunger gates P1 and P2, with gates B1 and B3 forming barriers to control the tunnelling between each dot and the reservoir. Gate B2 acts as the interdot barrier for controlling the tunnel rate within the double-dot via the tunnel coupling ($t_c$). A continuous AC microwave tone is applied on B2 for performing EDSR to extract the effective g-factor. 

Double-dot formation is confirmed by a periodic array of bias triangles in a charge stability map (Figure \ref{fig:deviceSchmatics}\textbf{C}) for a fixed source-drain bias of 2 mV. In Figure \ref{fig:deviceSchmatics}\textbf{C}, each bias triangle horizontal row represents a fixed dot occupancy for QD$_2$ while the occupation of QD$_1$ varies. Since the double-dot hole number is unknown, we focus on a single bias triangle row and vary the number of holes on QD$_1$ for a fixed QD$_2$ occupation, with the initial hole number denoted by (M,N). The bias triangle base marked by the blue star in Figure \ref{fig:deviceSchmatics}\textbf{D} shows Pauli spin-blockade for the lowest hole occupancy studied due to prohibited (odd, odd) to (even, even) hole transitions. Pauli spin-blockade can be lifted by spin-relaxation, producing a finite spin-leakage current ($I_{leak}$) as a result of various spin non-conserving processes. We observe an $I_{leak}$ value of 6 pA, as shown by the line-cut in Figure \ref{fig:deviceSchmatics}\textbf{E} (blue star) along the detuning axis of the bias triangle in Figure \ref{fig:deviceSchmatics}\textbf{D}. The spin-leakage current in Figure \ref{fig:deviceSchmatics}\textbf{E} includes contributions from SOI effects as well as spin-flip cotunneling, which must be separated in order to obtain information on the strength of the SOI from spin-orbit coupling.

\newpage 

\begin{figure} 
    \includegraphics[width=0.8\columnwidth]{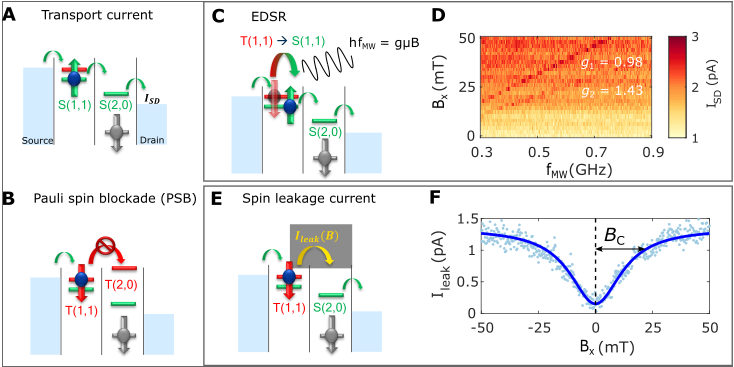}
	\caption{\label{fig:edsr_and_leakage} 1-D electric-dipole spin-resonance (EDSR) and spin-leakage current ($I_{leak}$) measurements.}  
 (\textbf{A}) Transport is allowed through the double-dot via the singlet ($S$) like states $S$(1,1) and $S$(0,2). 
 (\textbf{B}) Transport is prohibited via the triplet like ($T$) state $T$(1,1) to $S$(2,0) due to spin conservation and Pauli exclusion.  
 (\textbf{C}) A diagram showing how an AC electric-field frequency ($f_{MW}$) resonant with the Zeeman-splitting lifts Pauli spin-blockade via EDSR.
 (\textbf{D})$I_{SD}$ as a function of $B$ and $f_{MW}$ shows with two distinct resonant frequencies as a result of g-factors $g_1$ and $g_2$ from dots QD$_1$ and QD$_2$ respectively.
 (\textbf{E}) Pauli spin-blockade can also be lifted by the coupling between an applied B-field and the SOI as a result of S-T state mixing, producing a finite $I_{leak}$. 
 (\textbf{F}) $I_{leak}$ as a function of $B$ is fitted to a characteristic inverse Lorentzian to extract a critical magnetic-field $B_c$ = 12.6 mT, quantifying the relative spin-orbit strength.
\end{figure} 

To study the spin-orbit effects we operate our device in the (1,1)-(2,0) equivalent hole transition for Pauli spin-blockade, where (M,N) indicates the effective dot occupation. Double-dot transport is allowed between the singlet $S$(1,1)-$S$(2,0) like states (Figure \ref{fig:edsr_and_leakage}\textbf{A}) and blocked between the triplet and singlet like states $T$(1,1)-$S$(2,0) (Figure \ref{fig:edsr_and_leakage}\textbf{B}) owing to the Pauli-exclusion principle and energy inaccessibility of $T$(2,0). The diagram in Figure \ref{fig:edsr_and_leakage}\textbf{C} displays how EDSR is achieved by an AC microwave tone ($f_{MW}$) driving transitions between the $S$-$T$ states to lift spin-blockade according to $f_{MW}h = \boldsymbol{g}\mu_{B}\boldsymbol{B}$, where $\boldsymbol{g}$ is the hole g-factor, $\mu_{B}$ is the Bohr magneton and $\boldsymbol{B}$ is the magnetic-field along $x$ ($\boldsymbol{B_x}$). Figure \ref{fig:edsr_and_leakage}\textbf{D} shows $I_{SD}$ as a function of $f_{MW}$ and $\boldsymbol{B_x}$ for the spin-blocked region marked by the blue star in Figure \ref{fig:deviceSchmatics}\textbf{E}. Two distinct lines are observed in the colourmap where $I_{SD}$ is increased due to hole-spin resonance, yielding values of $g_1$=0.98 and $g_2$=1.43 from the line gradients. 

The coupling between $\boldsymbol{B}$ and spin-orbit coupling ($\boldsymbol{t_{SO}}$) can be probed by the magnetic-field dependence of $I_{leak}$ (shown schematically in Figure \ref{fig:edsr_and_leakage}\textbf{E}) since $I_{leak}$ represents the amount of SOI mixing between the $S$ and Zeeman-split $T$ states ($T_+$/$T_-$) \cite{PhysRevB.95.155416}. We measure $I_{leak}$ as a function of $B_x$ in the Figure \ref{fig:edsr_and_leakage}\textbf{F} line plot and observe a rapid increase in $I_{leak}$ with increasing magnetic-field. Information on the $\boldsymbol{B}$ dependence of $I_{leak}$ can also be extracted from the off-resonance $I_{SD}$ in Figure \ref{fig:edsr_and_leakage}\textbf{D}, however a current offset must be subtracted from $I_{SD}$ in Figure \ref{fig:edsr_and_leakage}\textbf{D} to obtain $I_{leak}$. For $B_x>$40 mT, $I_{leak}$ saturates as a product of optimally coupled $S$-$T_+/T_-$ states leading to a maximum spin-relaxation rate\cite{danon_pauli_2009}. This maximum spin-relaxation rate, or $I_{sat}$, is set by the reservoir/interdot tunnel rates and $\boldsymbol{t_{SO}}$. 
According to Figure \ref{fig:edsr_and_leakage}\textbf{F} nuclear hyperfine interactions are weak, since hyperfine mixing typically produce a characteristic sharp peak in $I_{leak}$ centered around $B$=0, which is not observed. The absence of strong hyperfine interactions is consistent with the atomic p-orbital nature of holes.

To quantify the spin-orbit mixing, we fit the experimental data in Figure \ref{fig:edsr_and_leakage}\textbf{F} to the expected form of $I_{leak}$ as a function of $\boldsymbol{B}$ in the Pauli spin-blockade regime
\begin{equation}
\label{eqn:1_leak}
I_{leak} = I_{sat} 
\left(1-\frac{8}{9}\frac{{B_c}^2}{\boldsymbol{B}^2 + B_c^2}\right)
+ I_B
\end{equation}
The saturation current due to spin-orbit mixing only is given by  $I_{sat}$, $I_B$ is the current offset at $B=0$ and $B_c$ is the critical magnetic-field that parameterises the spin-orbit mixing strength.
Based on this model, we extract $B_c$ = 12.6 mT for $B_x$, orientated along [$\bar{1}$10] (see Supplementary Section IV for fitting details). The orientation of $\boldsymbol{B}$ is important, since $B_c$ is highly dependent on any anisotropic effects in both the strength of spin-orbit mixing ($\boldsymbol{t_{SO}}$) and the dot g-tensor, which we discuss below.


\newpage

\begin{figure} 
    \includegraphics{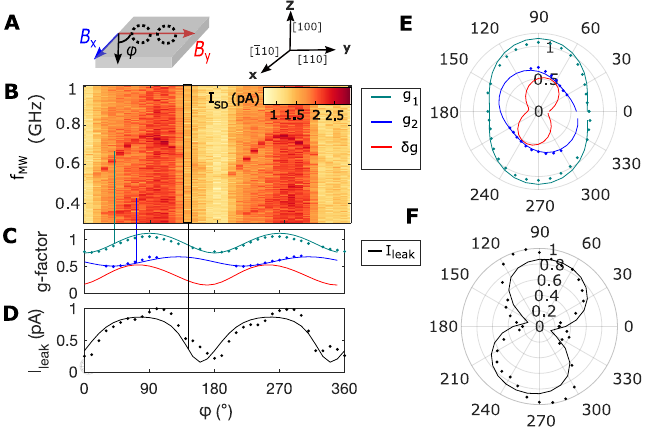}
	 \caption{\label{fig:vectorMag} Effective g-factor and $I_{leak}$ anisotropy in-plane ($y$$x$) for hole occupation number (M,N).}
  (\textbf{A}) Schematics depicting $\boldsymbol{B}$ rotation relative to crystal axes.(\textbf{B}) Electric-dipole spin-resonance (EDSR) as a function of $\phi$. $B_y$ is aligned along the transport direction ([110]), with $B_x$ perpendicular ([$\bar{1}$10]). (\textbf{C}) Extracted effective g-factors as a function of $\phi$, where $\delta \boldsymbol{g}$ is calculated via $|g_1 - g_2|$. (\textbf{D}) $I_{leak}$ is extracted from the average $I_{SD}$ for a given $\phi$. (\textbf{E}) To better visualize the anisotropies, polar plots are generated of $\boldsymbol{g_1}$, $\boldsymbol{g_2}$ and $\delta \boldsymbol{g}$, as well as $I_{leak}$ in (\textbf{F}) to compare. 
\end{figure} 
To study the effective g-factor and $I_{leak}$ anisotropy we measure $I_{SD}$ as a function of $\boldsymbol{B}$ orientation as shown schematically in Figure \ref{fig:vectorMag}\textbf{A} for an in-plane rotation. By performing EDSR, we are able to distinguish two hole-spin resonance lines in Figure \ref{fig:vectorMag}\textbf{B} as a function $f_{MW}$ and $\boldsymbol{B}$ angle ($\phi$) for a fixed $\boldsymbol{B}$ = 50 mT. Two lines are observed as a result of QD$_1$ and QD$_2$ possessing site-dependent g-factors, $g_1$ and $g_2$, which are extracted and plotted in Figure \ref{fig:vectorMag}\textbf{C}, along with the g-factor difference ($\delta \boldsymbol{g}$). $I_{leak}$ is obtained using the average $I_{SD}$ in Figure \ref{fig:vectorMag}\textbf{B} (minus an offset due to other spin-flip processes) and plotted in Figure \ref{fig:vectorMag}\textbf{D} as a function of $\phi$. To aid in visualizing the correlation between g-factor and $I_{leak}$ anisotropy, polar plots are generated of $\boldsymbol{g_1}$, $\boldsymbol{g_2}$ and $\delta \boldsymbol{g}$ in Figure \ref{fig:vectorMag}\textbf{D} and $I_{leak}$ in Figure \ref{fig:vectorMag}\textbf{E}. A 45$^{\circ}$ shift in the effective g-factor orientation is present between $\boldsymbol{g_1}$ and $\boldsymbol{g_2}$ in \ref{fig:vectorMag}\textbf{E}, attributed to variations in the local electronic environment surrounding each dot, together with strain and differences in confinement. Comparisons between $\delta \boldsymbol{g}$ and $I_{leak}$ anisotropy in Figure \ref{fig:vectorMag}\textbf{E} $\&$ \ref{fig:vectorMag}\textbf{F} yield a strong correlation, suggesting that $\delta \boldsymbol{g}$ contributes to spin-flip tunneling.

The mechanism linking $\delta \boldsymbol{g}$ and $I_{leak}$ has been reported as a coherent mixing of the non-polarized $T_0$ and $S(1,1)$ like states, where a larger $\delta \boldsymbol{g}$ results in faster $T_0$ spin-relaxation to $S$ as a consequence of the varying phase difference and enhanced mixing between the two states \cite{PhysRevLett.128.126803}. The correlation between the $\boldsymbol{t_{SO}}$ and $\delta \boldsymbol{g}$ maximum is believed to arise due to a large splitting between $T_{-/+}$-$S(0,2)$ and $T_0$-$S(1,1)$, and since the $S(1,1)-S(0,2)$ states are strongly coupled, faster spin-relaxation occurs between all $S$ and $T$ states. 

\newpage
\begin{figure} 
    \includegraphics[width=0.9\textwidth]{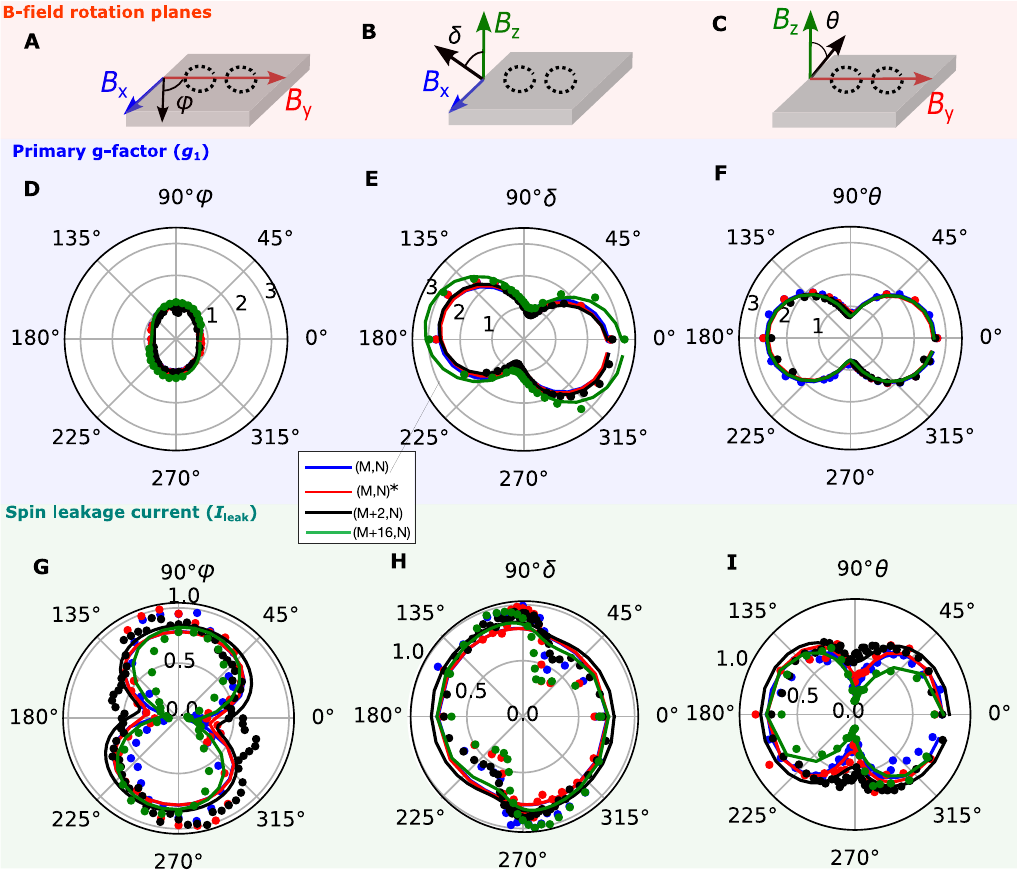}
	 \caption{\label{fig:3d_rotation}}  (\textbf{A}-\textbf{C}) Schematics defining $\boldsymbol{B}$ rotations in the $y$$x$, $z$$x$ and $z$$y$ planes. Polar plots of QD$_1$ effective g-factor ($\boldsymbol{g_1}$) (\textbf{D}-\textbf{F}) and $I_{leak}$ (\textbf{G}-\textbf{I}) for three different hole occupation numbers: (M, N)(blue), (M, N)* with different $t_c$ (red), (M+2, N) (black) and (M+16, N)(green). $g_1$ is extracted via EDSR, while $I_{leak}$ is measured using the off-resonance $I_{SD}$. The points and lines denote extracted data and fitting respectively. 
\end{figure} 

To investigate the impact of varying the hole number on the g-tensor and $\boldsymbol{t_{SO}}$ we rotate $\boldsymbol{B}$ in all three orthogonal planes $y$$x$, $z$$x$ and $z$$y$ (Figures \ref{fig:3d_rotation}\textbf{A}-\textbf{C}) to extract the effective g-factor $\boldsymbol{g_1}$ and $I_{leak}$ for three different dot occupations of QD$_1$. The anisotropy of $\boldsymbol{g_1}$ in Figures \ref{fig:3d_rotation}\textbf{D}-\textbf{F} is consistent at each dot occupation with very little variation in magnitude and orientation upon changing the number of holes. We also observe a similar insensitivity to changes in hole occupation number for $I_{leak}$ in Figures \ref{fig:3d_rotation}\textbf{G}-\textbf{I}.
Notably, when comparing the measured data for each plane in Figures \ref{fig:3d_rotation}\textbf{D}-\textbf{I}, $I_{leak}$ and $\boldsymbol{g_1}$ exhibit a similar dependence on the orientation of $\boldsymbol{B}$.

The calculated values of $t_c$ at each hole occupation are consistent with the extracted $B_c^{0}$, which are expected to be approximately proportional \cite{danon_pauli_2009}. The direction of $\boldsymbol{B}$ that resulted in the minimum $I_{leak}$ ($\boldsymbol{B} \parallel \boldsymbol{t_{SO}}$) corresponds to $\boldsymbol{t_{SO}}$ strongly aligned to the transport direction, which suggests that the surface Dresselhaus SOI is dominant, rather than Rashba SOI, since Rashba acts in-plane and perpendicular to transport\cite{rashba_spin_2005,kloeffel_direct_2018}. As such, choosing $\boldsymbol{B} \perp \boldsymbol{t_{SO}}$ would yield strong electrical driving of the qubit spin state,  while $\boldsymbol{B} \parallel \boldsymbol{t_{SO}}$ would improve coherence but limit control. In our double-dot, we assign the strong Dresselhaus interaction to variations in the inter-atomic spacing at the SiO$_2$ interface, breaking inversion symmetry and generating an electric-field which couples to the hole-spin in-plane.

One of the advantages of a g-tensor which is largely independent from the hole number and gate bias is that strain effects from electric-field gradients can be isolated. As such, from the small tilt where $\theta_g$$\neq0$ in Figure \ref{fig:3d_rotation}\textbf{E}, we assign the origin to non-uniform strain at the SiO$_2$ interface \cite{liles_electrical_2021}. The presence of non-uniform strain is further supported by the in-plane site-dependent g-factors of both dots shown in Figure \ref{fig:vectorMag}\textbf{E}, which modulate the g-tensor at each dot.


\newpage

In summary, the extracted g-tensor was found to be highly anisotropic, with a large out-of plane $\boldsymbol{g_1}$ $\approx$ 2.6 and a small in-plane $\boldsymbol{g_1}$ $\approx$ 0.8. A correlation between the g-tensor and spin-leakage current anisotropy was observed, which impacted the magnitude of $B_c$ and spin-orbit $S$-$T$ mixing. We determined the $\boldsymbol{t_{SO}}$ orientation to be aligned mostly parallel to transport, a characteristic of surface Dresselhaus SOI, in contrast to Rashba SOI that has largely been considered in hole dots. We also observed a strong correlation between $\delta \boldsymbol{g}$ and $I_{leak}$ as a result of mixing between $S$ and the non-polarized $T_0$ state. Comparisons between three different dot occupations yielded little to no-effect on the g-tensor and spin-relaxation, meaning the spin properties of our hole-dots are resilient to changes in hole number and gate bias.

\subsection{Experimental Methods}
The Si-MOS device consists of a three-layer gate structure with a sub-50 nm width and 5 nm isolation layer enables a tighter QD confinement potential. The measurements were performed in a BlueFors LD250 dilution refrigerator with a base temperature of 8 mK and an approximate electron temperature of 120 mK. Magnetic-field rotations to study the anisotropy in spin-leakage current and g-factor were achieved via a 6T-1T-1T vector magnet. The sample was mounted on a PCB with a LPF on the DC lines ($f_c =$ 130 kHz), and bias-tees on the RF lines (R=1.2 k$\Omega$, C = 1 nF) in order to apply microwave tones. Voltages for the DC were sourced through a Quantum Machines QDAC and an R$\&$S SGS100A to generate microwaves for EDSR. The spin-leakage was current converted into a voltage via a low-noise high-stability Basel instruments I-V converter with $f_c =$ 300 Hz, and measured by a Keithley K2000 multi-meter. 
\begin{acknowledgement}
 This work was funded by the Australian Research Council under projects LP200100019, DP200100147 and IL230100072. 
\end{acknowledgement}


\bibliography{UNSW_anisotropy_gtensor_sleak_draftV1}

\end{document}